# Non-Relativistic Electron Transport in Metals: A Monte Carlo Approach

Nima Ghal-eh, Farhad Rahimi, Mehrnoush Manouchehri

**Physics Department, School of Sciences, Ferdowsi University of Mashad, Iran**

**Abstract:** A simple Monte Carlo procedure is described for simulating the multiple scattering and absorption of unpolorized electrons with the incident energy in the range 1-50 keV moving through a slab of uniformly distributed material of given atomic number, density and thickness. The simulation is based on a screened Rutherford cross-section and Bethe continuous energy-loss equation. A FORTRAN program is written to determine backscattering, transmission and absorption coefficients, providing the user with a graphical output of the electron trajectories. The results of several simulations are presented by using various numbers of electrons, showing a good agreement with the experiment. The program is used to analyze the relation between the energy and the range of electron in the slab, the backscattering, absorption, transmission coefficients and the angular distribution.

# 1. Introduction
## to Monte Carlo Method

Monte Carlo is a numerical method for solving or simulating complex statistical problem [1]. In recent years when we are dealing with multi-parameter problems or those in which there is not an analytical solution, this method has been used frequently.

The main idea is to propose a statistical model, which is compatible with our problem or the problem is simulated exactly. Random parameters constructed are based on particular rules from which the phenomenon is sampling several times.

The advantage of this method is the ability of solving complex problems and the main disadvantage is having a huge number of computations and consequently a long run-time of computer program. For example to increase one digit to our result, the time consuming becomes more than hundred times.

Monte Carlo methods can theoretically reproduce any process where the interaction probabilities can be expressed statistically, such as interaction of electrons with matter. Using "random numbers", computers are able to create a statistical history for each particle. Then, an individual particle can experience several scattering interactions before absorption or leakage from the system. The "random" number is between 0 and 1, and it is used in all interactions to determine which kind of interactions (absorption, elastic scattering, etc) takes place, how much energy is lost, what the direction of the particle (in scattering) is, etc. In this section, a brief basic description of the Monte Carlo methods will be given. The following description gives the method of sampling of a collision along a track.

The "free path length", L, of an electron (i.e. distance between successive collisions) is a random variable. The probability, P(L)dL, of occurrence of a collision between L and L+dL along its path is:

$$P(L)dL = e^{-\Sigma_t L} \Sigma_t \, dL \qquad (1\text{-}1)$$

where

$$\Sigma_t = \sum_i^n \sigma_i N_i = \sum_i^n \Sigma_i \qquad (1\text{-}2)$$

and $\Sigma_t$ total macroscopic cross section, representing the probability of the interaction in the unit length of the particle path, $N_i$, is number of atoms per unit volume and $\sigma_i$ is the differential cross section. Suppose a random number $\xi$ in the interval [0,1]. It is introduced as:

$$\xi = \int_0^L e^{-\Sigma_t L} \Sigma_t \, dL = 1 - e^{\Sigma_t L} \qquad (1\text{-}3)$$

It then follows that

$$L = -\frac{1}{\Sigma_t} \ln(1 - \xi) \qquad (1\text{-}4)$$

But, because $1-\xi$ is distributed in the same manner as $\xi$, it may be replaced by $\xi$. Then we obtain a well-known expression for the distance between collisions.

$$L = -\frac{1}{\Sigma_t} \ln(\xi) \qquad (1\text{-}5)$$

The total macroscopic cross section is $\Sigma_t = \Sigma_a + \Sigma_s$, where $\Sigma_a$ and $\Sigma_s$ are the macroscopic absorption and scattering cross sections, respectively; then $\Sigma_s/\Sigma_t$ indicates the probability of scattering while $\Sigma_a/\Sigma_t$ gives the probability of absorption. Now, having the random number $\xi$, the type of interaction the particles undergoes can be determined, if $\Sigma_a/\Sigma_t < \xi < 1$ an absorption interaction can be supposed to take place and if $0 > \xi > \Sigma_a/\Sigma_t$ then a scattering interaction occurs.

In order to determined the interacting atom, with the random number $\xi$ within interval [0,1], the $k^{th}$ atom is chosen as the collision atom, iin this case $P_{k-1} < \xi \leq P_k$ where $P_k = \Sigma_k / \Sigma_{tot}$. Angles and the energies of the particles after collisions are determined similarly. The cosine of the scattering angle is sampled from energy dependent angular distribution formula for each collision atom.

## 2. Physics

The fundamental force, which describes the interaction between the incident electrons and target particles, is the Coulomb force. The ways of losing energy for an electron are ionization, Compton scattering, pair production, bremsstrahlung, etc. The energy range of interest determines the loss mechanisms and scattering processes. In this paper we consider electrons with 1-50 keV energy, and the ionization is the dominant energy loss mechanism and elastic collision with nuclei produce the majority of the relatively large angular deflections.

The essential factors, which we need to solve the problem, are: a) Determination of energy lost during scattering event. b) Differential scattering cross-section, which characterizes the scattering process. c) Determination of the distance between scattering events.

There is an expression, derived by H. Bethe, which gives the kinetic energy lost by a nonrelativistic electron as it passes through the matter of length ds [2].

$$\frac{dT}{ds} = -\left(\frac{2\pi e^4}{T}\right) NZ \ln\left(\frac{2T(eV)}{11.5Z}\right) \quad (2\text{-}1)$$

In above equation, T is the kinetic energy of the electron, e is the electron charge, N is the number of target atoms/cm$^3$, and Z is the atomic number. Eq. (2-1) can be written in terms of keV/μm, atomic weight and density of the target material:

$$\frac{dT}{ds} = -7.83\left(\frac{\rho Z}{AT}\right)\ln\left(\frac{174T}{Z}\right)\left(\frac{keV}{\mu m}\right) \quad (2\text{-}2)$$

In which, ρ(g/cm$^3$) is density of target, A(g) is atomic weight of the target, and T(keV) is electronic kinetic energy.

Here we use the Rutherford differential cross-section to describe the elastic nuclear scattering. Due to infinite range of Coulomb force, all incident particles will scatter to some extent even if their impact parameter is arbitrary large. So the total Rutherford cross-section is infinite as well. To avoid this difficulty we use a so-called shielding technique. We assume its surrounding electrons shield the target nucleus, and as a result the Coulomb field is decreased exponentially.

$$\frac{d\sigma}{d\Omega} = \left(\frac{Z(Z+1)e^4}{p^2\upsilon^2}\right)\left(\frac{1}{(1-\cos\theta+2\beta)}\right) \quad (2\text{-}3)$$

Where $\beta = 0.25\left(\frac{1.12\lambda_0\hbar}{p}\right)$, $\lambda_0 = \frac{Z^{1/3}}{0.885a_0}$,

p=mυ (the electron momentum), and $a_0$=Bohr radius. The angle θ represents the scattering angle. The total cross-section is:

$$\sigma_T = \iint d\Omega \frac{d\sigma}{d\Omega} = \left(\frac{\pi}{\beta(\beta+1)}\right)\left(\frac{Z(Z+1)e^4}{p^2\upsilon^2}\right) \quad (2\text{-}4)$$

If we assume that the attenuation has an exponential form, the mean free-path of electron is given by:

$$\lambda = \int s\exp[(-\rho N_0 \sigma_{T/A})s]ds \int \exp[(-\rho N_0 \sigma_{T/A})s]ds$$

$$= \frac{1.02\beta(1+\beta)AT^2}{Z(Z+1)\rho} \quad (2\text{-}5)$$

$N_A$ is Avogadro's number.

## 3. Geometry

A typical electron path is illustrated in fig.1. The extent of its redirection is determined by the magnitudes of the scattering angle θ and the azimuthal angle ϕ. As electron moves some characteristic distances in a given direction it encounters a scattering center.

The scattering deflects the electron by an angle θ relative to its previous direction, which is governed by differential cross-section. The azimuthal angle ϕ is assumed to be a uniformly distributed random angle between 0 and 2π rad in each scattering.

There are two important frames of reference: 1) Laboratory frame, which is rigidly attached to the slab. 2) Scattering frame, which co-moves with the electron.

Questions as how many electrons are transmitted, how many are backscattered, and how many are absorbed by the material are considered in laboratory frame of reference. Whereas the questions concerning how many electrons cross various material boundaries are simple to be considered in the scattering frame.

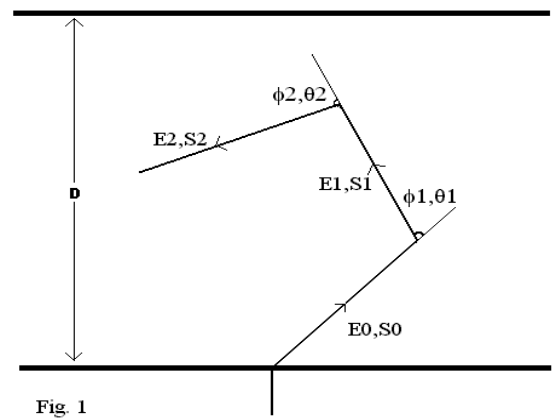

Fig. 1

Now, we try to represent the connection between these two frames of reference. As illustrated in fig. 2, the direction of the electron is defined by a unit vector $\mathbf{v}_n$ specified by $\theta_n$, $\phi_n$ relative to axes fixed in the laboratory frame. The scattering angles $\theta$ and $\phi$ are defined in the scattering frame which uses $\mathbf{v}_n$ to define $\mathbf{z}'$ axis. The electron travels in the direction $\mathbf{v}_n$ until it undergoes a scattering defined by the scattering angle $\theta$. The new direction of the electron in the laboratory frame is defined by the unit vector $\mathbf{v}_{n+1}$. Our task, now, is to find $\theta_{n+1}$ and $\phi_n$ in terms of $\theta_n$, $\phi_n$, $\theta$, and $\phi$.

According to fig.2, we have:
$\mathbf{v}_n = \sin\theta_n \cos\phi_n \mathbf{i} + \sin\theta_n \sin\phi_n \mathbf{j} + \cos\theta_n \mathbf{k}$
(3-1a)
or
$\mathbf{v}_{n+1} = \sin\theta_{n+1} \cos\phi_{n+1} \mathbf{i} + \sin\theta_{n+1} \sin\phi_{n+1} \mathbf{j} + \cos\theta_{n+1} \mathbf{k}$ (3-1b)
$\mathbf{v}_{n+1} = \sin\theta \cos\phi \mathbf{i}' + \sin\theta \sin\phi \mathbf{j}' + \cos\theta \mathbf{k}'$
(3-1c)
where $\mathbf{i}'$, $\mathbf{j}'$, and $\mathbf{k}'$ refer to scattering frame of reference. We have:
$\mathbf{k}' = \mathbf{v}_n$
$\mathbf{j}' = \mathbf{v}_n \times \mathbf{k} / - \mathbf{v}_n \times \mathbf{k}- = \mathbf{v}_n \times \mathbf{k} / \sin\theta$
$= \sin\theta_n \mathbf{i} - \sin\phi_n \mathbf{j}$
$\mathbf{i}' = \mathbf{j}' \times \mathbf{k}' = (\mathbf{k} - \cos\theta_n \mathbf{v}_n) / \sin\theta$
(3-2)
or
$\mathbf{v}_{n+1} \cdot \mathbf{k} = \cos\theta_{n+1} = \cos\theta_n \cos\theta + \sin\theta_n \sin\theta \cos\phi$

$\mathbf{v}_{n+1} \cdot \mathbf{v}_n = \cos\theta$
$= \cos\theta_n \cos\theta_{n+1} + \sin\theta_n \sin\theta_{n+1} \cos(\phi_{n+1} - \phi_n)$

$\cos(\phi_{n+1} - \phi_n)$
$= (\cos\theta - \cos\theta_n \cos\theta_{n+1}) / \sin\theta_n \sin\theta_{n+1}$
(3-3)

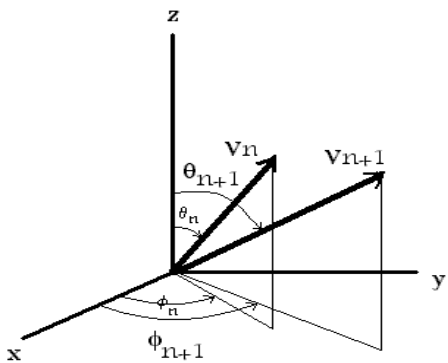

Fig. 2

## 4. Simulation and Programming

The initial electron's direction of motion is given by $\mathbf{v}_n$ ($\theta_n$, $\phi_n$) and its kinetic energy is $T_n$. An elastic scattering defined by angles $\theta$ and $\phi$ causes electron to move in the direction given by $\mathbf{v}_{n+1}$ ($\theta_{n+1}$, $\phi_{n+1}$) [3].

The probability for finding an electron scattered through an angle $\theta$ is given by:
$$P(\theta) = \frac{1}{\sigma_T} \iint d\Omega \left(\frac{d\sigma}{d\Omega}\right) = \frac{(1+\beta)(1-\cos\theta)}{(1+2\beta-\cos\theta)}$$
(4-1)
or
$$\cos\theta = 1 - \frac{2\beta P(\theta)}{1 + \beta - P(\theta)}$$
(4-2)

Since we consider unpolorized electrons, there is no other direction in space to fix the angle $\phi$, which means any angle between 0 and $2\pi$ is accessible. So by using two random numbers $R_\theta$ and $R_\phi$, the angles $\theta$ and $\phi$ can be easily derived.
$$\cos\theta = 1 - \frac{2\beta R_\theta}{1 + \beta - R_\theta}$$

$\phi = 2\pi R_\phi$ (4-3)

If N(0) electrons travel a path length s, the remaining number of electron will be:
N(s) = N(0) exp(–s/$\lambda$) (4-4)
where $\lambda$ is the mean free-path of the electron for an exponential attenuation. The probability for an electron to interact after path length s is:
P(s) = [N(0) – N(s)] / N(0)
= 1 – exp(–s/$\lambda$) (4-5)
or,
s = – $\lambda$ ln [1-P(s)] (4-6)
Since p(s) is between 0 and 1, so [1-p(s)] also varies between 0 and 1. It means that by choosing a random number $R_s$, we can find the path length traveled by the electron. The energy loss in each collision is then (dT/ds)s.

## 5. Conclusion

As illustrated in the figures below, the FORTRAN program output is shown in terms of different initial angles, different energies and different target metal. In the last figure, we have used a three-layer slab from which we can investigate the electron transition mechanisms through soft tissues of the body [4,5]. Our results are in good agreement with others' experimental data [6,7].

## 6. References

[1] S Weinzierl, hep-ph/0006269
[2] H A Bethe, Ann.Phys.5, **325**, (1930)
[3] S Jadach, physics/9906056
[4] G S Sidhu, et. al, Radiat. Prot. Dosim. **86**(3), pp 207-216 (1999)
[5] B V Thirumala et. al, Med.Phys. **12**(6), 1985
[6] W Williamson, G C Duncan, Am.J.Phys, **54**(3),1986
[7] K Murata, J.App.Phys,**45**,4110 (1974)

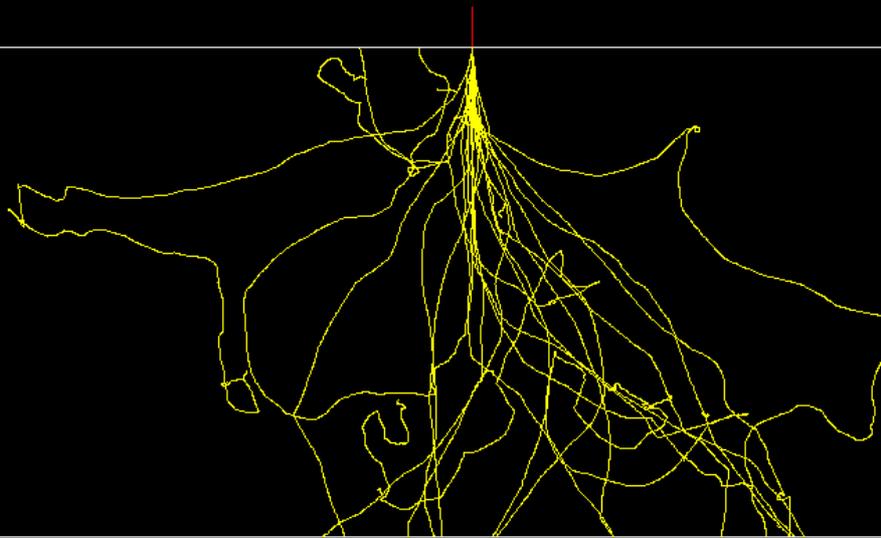
number of electrons = 20, energy= 20keV, gold

figure 1

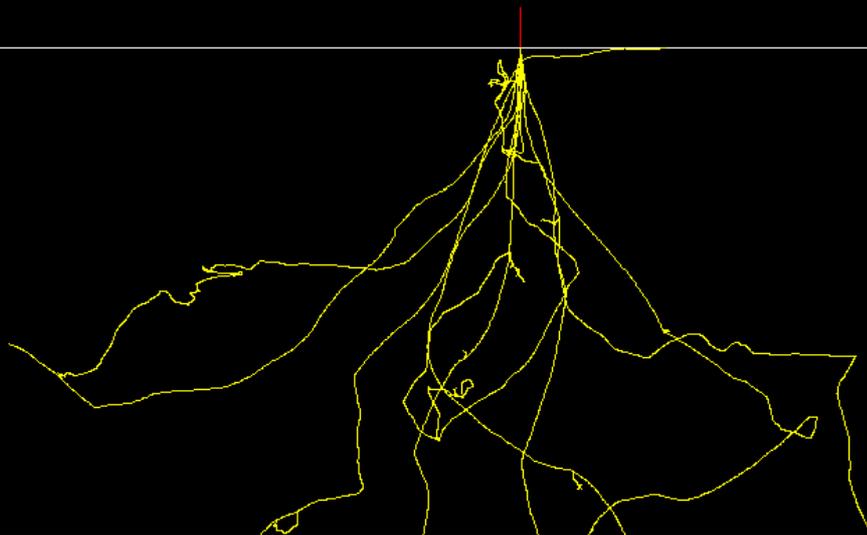
number of electrons=10, energy=20keV, gold

figure 2

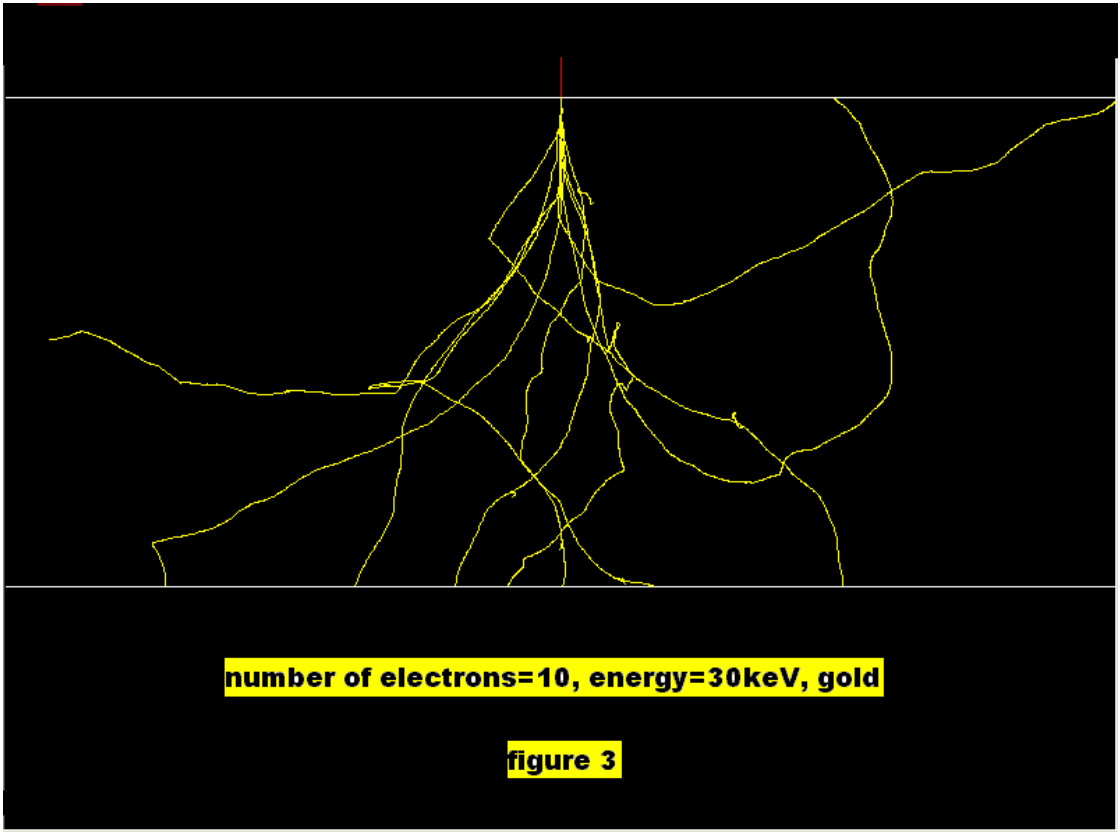

number of electrons=10, energy=30keV, gold

figure 3

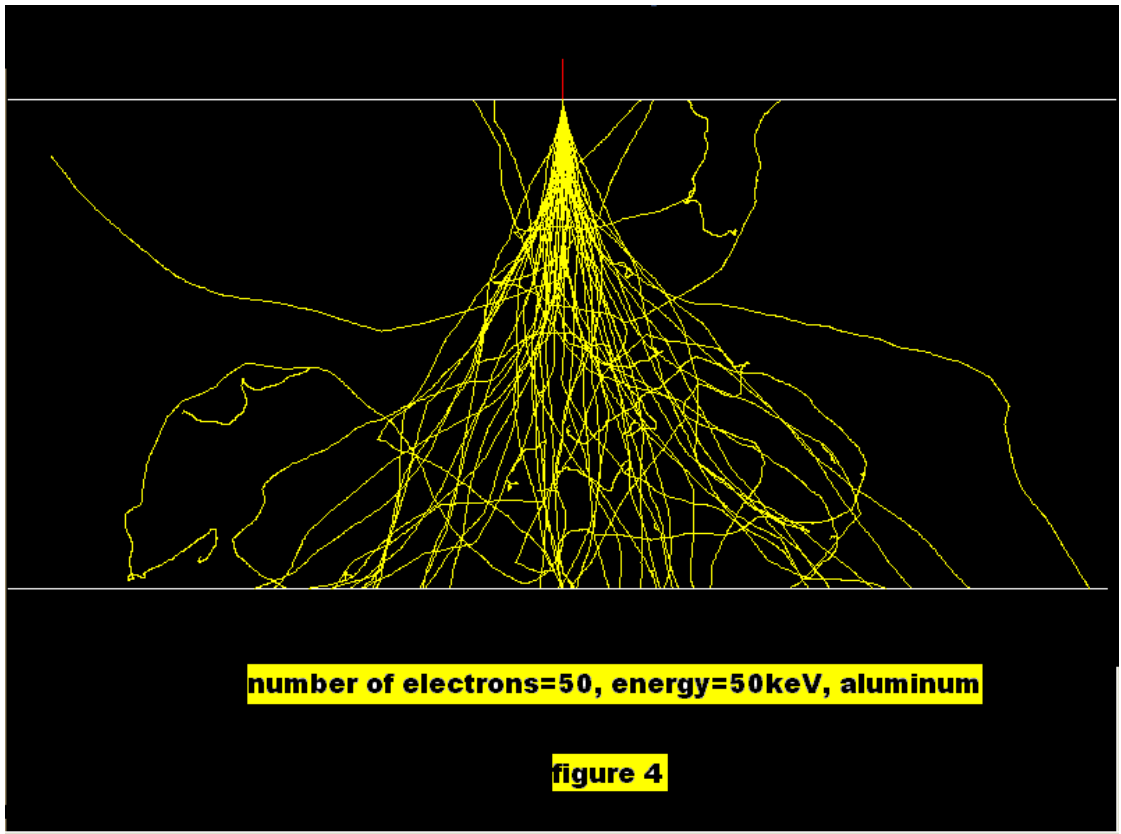

number of electrons=50, energy=50keV, aluminum

figure 4

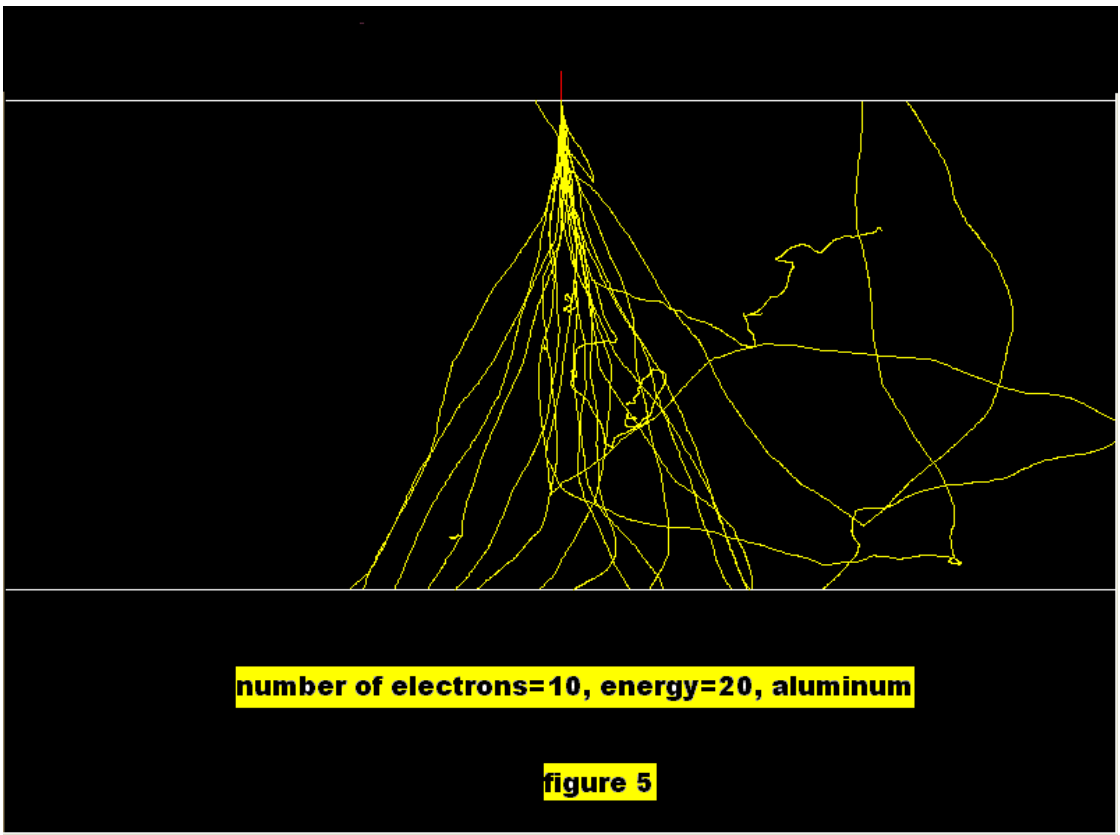
number of electrons=10, energy=20, aluminum

figure 5

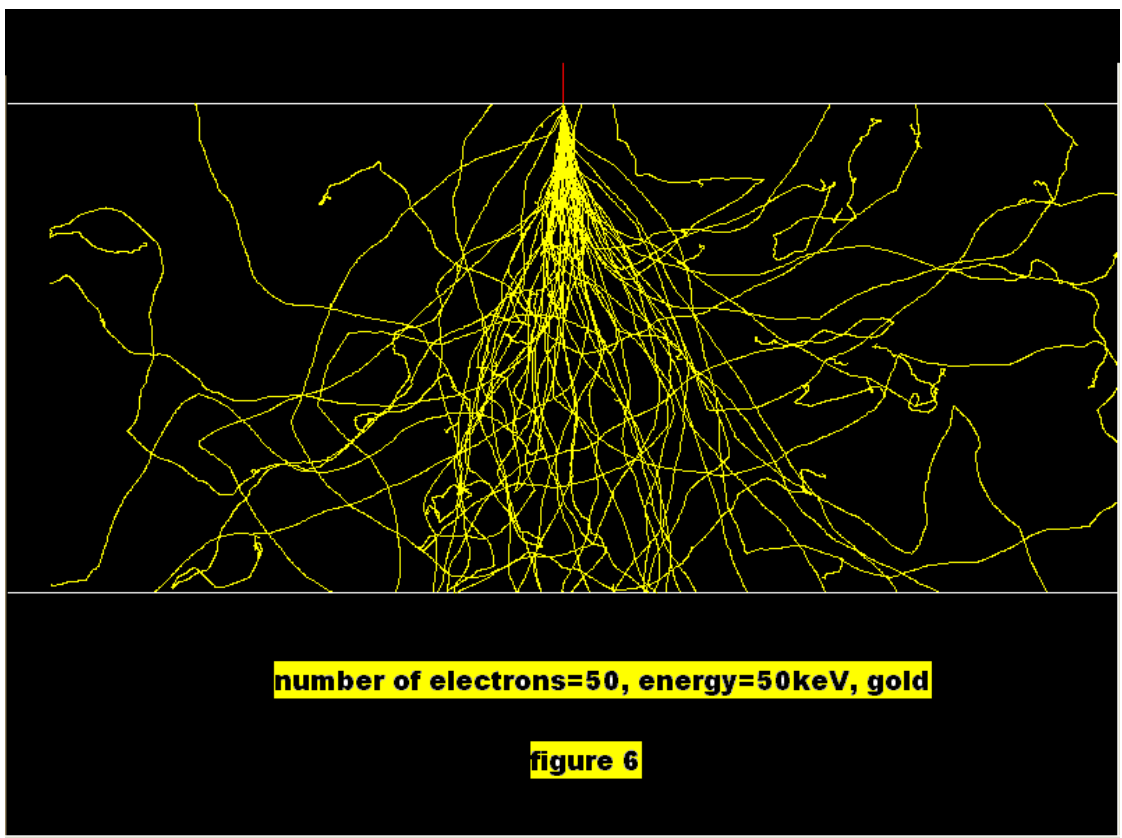
number of electrons=50, energy=50keV, gold

figure 6